\documentclass[twocolumn,aps,prd,epsf]{revtex4}
\usepackage{epsfig}
\usepackage{color}
\usepackage{amsmath}

\begin{document}

\title{ 
First study of the three-gluon static potential in Lattice QCD
}
\author{
M. Cardoso}
\author{
P. Bicudo}
\affiliation{CFTP, Departamento de F\'{\i}sica, Instituto Superior T\'ecnico,
Av. Rovisco Pais, 1049-001 Lisboa, Portugal}

\begin{abstract}

We estimate the potential energy for a system of three static gluons in Lattice QCD.
This is relevant for the different models of three-body glueballs have been proposed in the 
literature, either for gluons with a constituent mass, or for massless ones. 
A Wilson loop adequate to the static hybrid three-body system is developed. 
We study different spacial geometries, to compare the starfish model with the
triangle model, for the three-gluon potential. We also study two different colour structures, 
symmetric and antisymmetric, and compare the respective static potentials. 
A first simulation is performed in a $24^3 \times 48$ periodic Lattice, with $\beta=6.2$ 
and $a \sim 0.072$ fm. 
\end{abstract}

\maketitle

\section{Introduction}

We explore, in Lattice QCD, the static potential of the three-body glueball system composed
of three gluons, using Wilson loops. 
The interest in three-body gluon-gluon-gluon systems is increasing 
in anticipation to the future experiments BESIII at IHEP in Beijin, GLUEX at 
JLab and PANDA at GSI in Darmstadt, dedicated to study the mass range of 
charmonium, with a focus in its plausible hybrid excitations. Even before the
glueballs are discovered, the study of two-gluon and three--gluon glueballs
are respectively relevant to the pomeron 
\cite{LlanesEstrada:2000jw,Meyer:2004jc}
and to the odderon
\cite{LlanesEstrada:2005jf}. 
Thus several models of three-gluon models have already started to be developed
\cite{LlanesEstrada:2005jf,Hou:1982dy,Mathieu:2006bp,Buisseret:2007hf,Boulanger:2008aj,Mathieu:2008pb}.

The relevance of computing the static potentials in Lattice QCD for 3-gluon models is partly motivated by the 
plausible existence of a constituent mass for the gluon. 
Several evidences of a gluon effective mass of 600-1000 MeV,
much larger than $\Lambda_{QCD}$,  exist
from the Lattice QCD gluon propagator in Landau gauge, 
\cite{Leinweber:1998uu,Oliveira_propagator}, 
from Schwinger-Dyson and Bogoliubov-Valatin solutions for the gluon propagator in
Landau gauge 
\cite{Fischer:2002eq},
from the analogy of confinement in QCD to 
supercondutivity 
\cite{Nielsen:1973cs},
from the Lattice QCD breaking of the adjoint string 
\cite{Michael:1985ne},
from the Lattice QCD gluonic 
excitations of the fundamental string 
\cite{Griffiths:1983ah}
from constituent gluon models 
\cite{Hou:1982dy,Szczepaniak:1995cw,Abreu:2005uw}
compatible with the Lattice QCD glueball spectra 
\cite{glulat1,glulat2,glulat3,glulat4}, 
and with the Pomeron trajectory for high energy scattering 
\cite{LlanesEstrada:2000jw,Meyer:2004jc}. 
Furthermore, even for modelling massless gluons, the knowledge of
a static potential would at least provide one of the components
of the dynamical potential. For instance, the static quark-antiquark potential is 
frequently applied to light quarks. 

The Wilson loop method was devised to extract, from pure-gauge QCD, the static potential 
for constituent quarks and to provide a detailed information on the confinement 
in QCD. In what concerns gluon interactions, the first Lattice studies were
performed by Michael 
\cite{Michael:1985ne,Campbell:1985kp}
and Bali extended them to other SU(3) representations 
\cite{Bali:2000un}.
Recently Okiharu and colleagues 
\cite{Okiharu:2004ve,Okiharu:2004wy}
studied for the first time another class of exotic hadrons, extending the Wilson loop of three-quark baryons
to tetraquarks and to pentaquarks. Very recently, Bicudo, Cardoso and Oliveira  continued
the Lattice QCD mapping of the static potentials for exotic hadrons, with the
study of the hybrid quark-antiquark-gluon static potential,
\cite{Bicudo:2007xp,Cardoso:2007dc}.

\begin{figure}[t!]
\begin{picture}(350,90)(0,0)
\put(-20,-20){\includegraphics[width=0.7\textwidth]{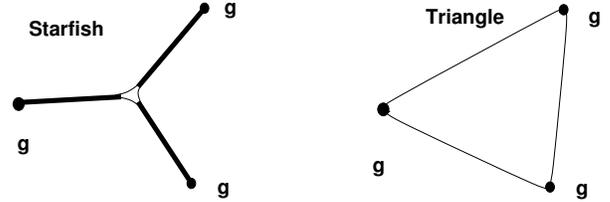}}
\end{picture}
\caption{The starfish-like  and triangle-like possible geometries for the strings
in the static three-gluon system.} 
\label{starfish}
\end{figure}

In this paper we study the three-gluon potentials in Lattice QCD. We address two novel and important questions.
Noticing that with three gluons two different colour singlets can be constructed, symmetric or antisymmetric,
we study whether the respective interactions are identical or different. This will be further detailed in Section II.
Moreover, noticing that a gluon may couple to one adjoint string, or to a pair of fundamental strings, we study 
whether the potential is amenable to a triangle-shaped triplet of fundamental strings or to a starfish-shaped triplet 
of adjoint strings, as depicted in  Fig. \ref{starfish}. A similar discussion on the shape of the baryonic strings has
been addressed in Lattice QCD 
\cite{Jahn:2003uz,deForcrand:2005vv,Takahashi:2002it,Suganuma:2003cv,Suganuma:2004hu,Takahashi:2004rw,Yamamoto:2007nn,Suganuma:2008ej,Yamamoto:2008fm,SilvestreBrac:2003sa}

In  particular, our study of the hybrid system already indicated
\cite{Bicudo:2007xp,Cardoso:2007dc}
that it would be interesting  
to study three-body glueballs, relevant for the odderon problem
\cite{LlanesEstrada:2005jf}. 
 Notice that in Lattice QCD, using the adjoint representation of SU(3), Bali 
\cite{Bali:2000un} 
found that the adjoint string is compatible with
the Casimir scaling, where the Casimir invariant $\lambda_i \cdot \lambda_j$ 
produces for the $gg$ interaction a factor $9/4$ times larger than the $q \bar q$ interaction. 
With three gluons, a triangle formed by three fundamental strings might  costs less energy
than three adjoint strings with a starfish-like geometry, depicted in Fig. \ref{starfish}.
The three-gluon potential may be similar to a sum 
of three mesonic-like quark-antiquark interactions, plus a repulsion acting only when 
there is superposition of the fundamental strings. This question is also related to the 
superconductor (Type-I versus Type-II) model  for confinement, where
 flux tubes repel each other in Type-II superconductors, while  in Type-I superconductors 
they attract each other and tend to fuse in excited vortices
\cite{bookGennes}. 
A first evidence of QCD string repulsion was indeed found in our very recent study of the hybrid potential
\cite{Bicudo:2007xp,Cardoso:2007dc}.
The understanding of the three-gluon potential in 3+1 dimension Lattice QCD
will further clarify our understanding of  confinement.

In Section II we derive a class of Wilson Loops adequate to study the
static hybrid potential. This paper is mainly analytical, and in Section III we 
discuss theoretically the important questions of the best Wilson Loops to 
distinguish the triangle from the starfish string groundstates, and of  the 
differences of the symmetric to antisymmetric potentials. 
In Section IV we present the first results of our numerical
Monte-Carlo simulations, and conclude.

\section{Three Gluon Wilson Loop   }

We first construct a wavefunction with three gluons. This wavefunction will be the
starting point of the Wilson Loop. Due to confinement, a hadron, system composed 
of quarks, antiquarks or gluons, must be a colour singlet. 

Each gluon is a state of the adjoint, or octet $\mathbf{8}$, representation of $SU(3)$. 
With the tensor product of two gluons, different representations of $SU(3)$ can be 
constructed,
\begin{equation}
\mathbf{8} \otimes \mathbf{8} = \mathbf{1} \oplus \mathbf{8} \oplus \mathbf{8} \oplus \mathbf{10} \oplus
\mathbf{10} \oplus \mathbf{27}
\label{twooctet}
\end{equation}
including a singlet  $\mathbf{1}$ and two octets  $ \mathbf{8}$. When we couple three gluons, we get 
not just one colour singlet, but two colour singlets (plus many other representations), resulting from coupling this 
third octet to each of the two octets in the right hand side of  eq. (\ref{twooctet}),
\begin{equation}
\mathbf{8} \otimes \mathbf{8} \otimes \mathbf{8} =
\mathbf{1} \oplus \mathbf{1} \oplus \mathbf{8}\oplus \cdots
\label{threeoctet}
\end{equation}
To arrive at the wavefunction for the two colour singlets, it is sufficient to study the product of two Gell-Mann matrices,
since it already produces the relevant colour singlet and colour octets resulting from  eq. (\ref{twooctet}),
\begin{equation}
\lambda^a  \lambda^b = \frac{2}{3} \delta^{ab} + i f_{abc} \lambda^c + d_{abc} \lambda^c
\end{equation}
and thus the product of three Gell-Mann matrices already produces the two possible colour singlets,
that we single out in the trace of the product of the three Gell-Mann matrices,
\begin{equation}
tr \left\{ \lambda^a \lambda^b  \lambda^c  \right\} =   2  i   f_{abc}+ 2 d_{abc}\   ,
\label{trace}
\end{equation}
and thus the two possible color singlet wavefunctions of three gluons are,
\begin{eqnarray}
	| \Psi^A \rangle &=& f_{abc} | a b c \rangle \ ,
\nonumber \\
	| \Psi^S \rangle &=& d_{abc} | a b c \rangle \ ,
\label{wavefunctions}
\end{eqnarray}
where the first combination is anti-symmetric and the second is symmetric with respect to the exchange of two gluons.

We build the three-gluon Wilson loop operator inspired in the three-quark case of the baryon. 
In the baryon we have a colour singlet wavefunction given by
\begin{equation}
	| \Psi^{Baryon} \rangle = \epsilon_{ijk} | i j k \rangle \ ,
\label{3qwf}
\end{equation}
and the corresponding Wilson loop is
\begin{equation}
	W_{3q} =  \epsilon_{ijk} \epsilon_{i'j'k'} X^{ii'} Y^{jj'} Z^{kk'}\  ,
\label{3qlooppaths}
\end{equation}
Where $X$, $Y$ and $Z$ are the elementary paths of the three quarks, each composed of the product of successive elementary links 
$U$ starting and ending in wavefunctions of the form (\ref{3qwf}).

\begin{figure}[t!]
\begin{picture}(350,120)(0,0)
\put(-30,-10){\includegraphics[width=0.6\textwidth]{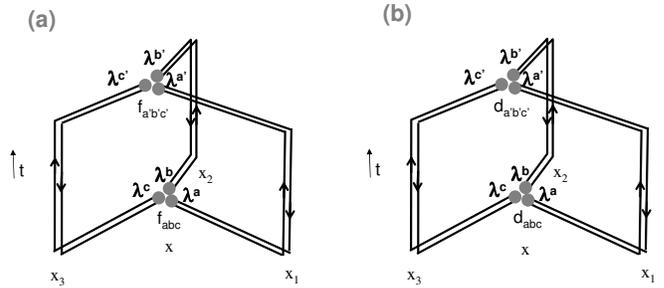}}
\end{picture}
\caption{Wilsons loop for the $g \, g \, g$ potential, (a) for the symmetric
colour wavefunction and (b) for the antisymmetric colour wavefunction} 
\label{Wilson loop diagram}
\end{figure}

In the three-gluon-glueball case we proceed similarly, developing adjoint paths $\widetilde X$, $\widetilde Y$ and $\widetilde Z$  starting 
either from the symmetric, or antisymmetric, colour singlet 
wavefunctions(\ref{wavefunctions}), as illustrated in Fig. \ref{Wilson loop diagram}.
Each adjoint path is composed of the product of successive adjoint links, 
corresponding to gluons, composed of matrices of the $SU(3)$ adjoint or octet representation, 
given in terms of the fundamental representation  ones by using the formula
\begin{equation}
{	\widetilde U_\mu (x)}^{ab} = \frac{1}{2} Tr \left\{ \lambda^a U_\mu (x) \lambda^b  \left[U_\mu (x)\right]^{\dagger} \right\}  \  .
\end{equation}

\begin{figure}[t!]
\begin{picture}(350,60)(0,0)
\put(0,-40){\includegraphics[width=0.6\textwidth]{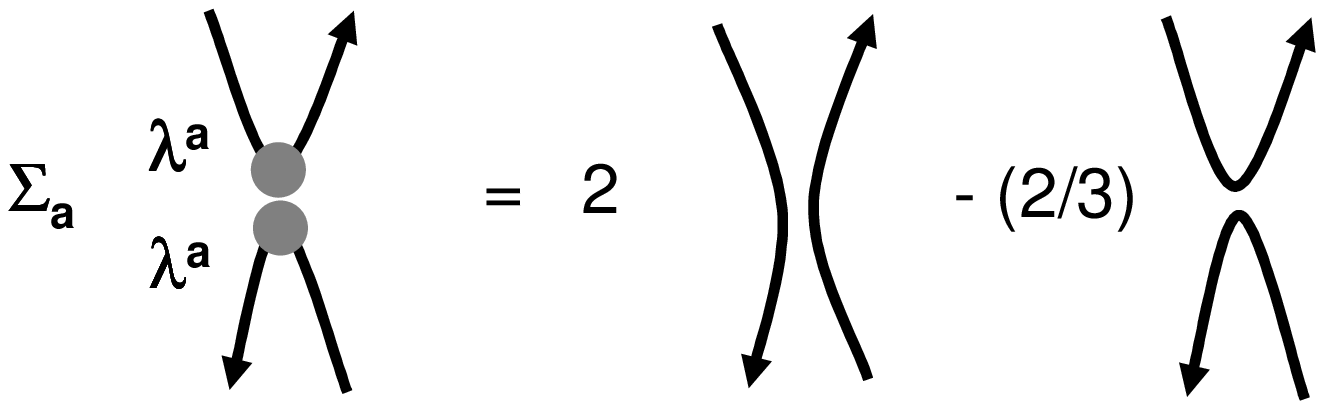}}
\end{picture}
\caption{Graphical version of the Fierz relation, showing that when two disconnect
paths touch each other at the same point where a pair of Gell-Mann matrices is 
summed in their indices, this is equivalent to connecting the paths in two different ways, 
both gauge invariant. }
\label{FierzContraction}
\end{figure}

Notice that these adjoint links are unitary matrices, as expected by a representation
of $SU(3)$ ,
\begin{eqnarray}
{\sum_b	}  \ {\widetilde U}^{ab}\  {{\widetilde U}^ \dagger }{}^{ bc }
&=&
 \sum_b	
 \frac{1}{4} Tr [ U^{\dagger}\lambda^a U \lambda^b  ]  
 Tr[ \lambda^b U^{\dagger} \lambda^c  U ]  
 \nonumber \\ &=&
  {1 \over  2} Tr [ U^{\dagger}\lambda^a U  U^{\dagger} \lambda^c  U ]  
 \nonumber \\ 
&& - {1 \over 6} Tr[ U^{\dagger}\lambda^a U ]  Tr[ U^{\dagger} \lambda^c  U ]  
  \nonumber \\ &=&
 \delta^{ac} \ ,
\end{eqnarray}
where we used the Fierz relation,
\begin{equation}
\sum_a \lambda^a_{ij} \lambda^a_{kl} = 2 \left( \delta_{il}\delta_{jk} - \frac{1}{3} \delta_{ij}\delta_{kl} \right)
\label{Fierz relation}
\end{equation}
illustrated in Fig. \ref{FierzContraction}, to contract the $\lambda^b$ matrices.

We  now explicitly derive the operator for the three-gluon Wilson loop. 
In the limit of arbitrarily large gluon masses, a non-relativistic potential $V$ can be
derived from the large time behaviour of euclidean time propagators. 
Typically, one has a meson operator $\mathcal{O}$ and computes the Green
function,
\begin{equation}
   \langle 0| \, \mathcal{O} (t) \, \mathcal{O} (0) \, | 0 \rangle ~
  \longrightarrow ~  \exp \{ - V t \}
\label{Green function}
\end{equation}
for large $t$. Different types of operators allow the definition of different
potentials. 
We can construct the three-gluon Wilson loop starting from the gluonic operator, 
\begin{eqnarray}
\mathcal{O}_{3g}^A (x) &=&  \,  f_{abc} \, 
 \Big[  g^a(x) \Big]  \,  \Big[ g^b(x) \Big]  \,  \Big[ g^c(x) \Big]    \, ,
\label{mesonmesonmeson}
\end{eqnarray} 
where the second operator $\mathcal{O}^S$  is constructed  replacing $f_{abc}$ by $d_{abc}$. 
In eq. (\ref{mesonmesonmeson}) the three
octets are situated in the same point $x$.  Using the Lattice links to comply with 
gauge invariance, the second operator in eq. (\ref{mesonmesonmeson})  can be made 
non-local to separate the three octet operators,
\begin{eqnarray}
& \mathcal{O}_{3g}^A  & \hspace{-.15 cm}(x, x1,x2,x3) =  f_{abc}
\label{op888}
\\
&&
  \Big[  \widetilde U_{\mu_1} (x) \cdots  
\widetilde U_{\mu_1} (x + (r_1-1)\hat{\mu}_1) {\Big]^a}_{a_1}          g ^{a_1}(x + r_1 \hat{\mu}_1 )
\nonumber \\
&&   
  \Big[  U_{\mu_2} (x) \cdots 
U_{\mu_2} (x + (r_2-1)\hat{\mu}_2) {\Big] ^b}_{b_1}         g ^{b_1}(x + r_1 \hat{\mu}_2) 
\nonumber \\
&&  
  \Big[   U_{\mu_3} (x) \cdots 
  U_{\mu_3} (x + (r_3-1)\hat{\mu}_3) {\Big]^c}_{c_1}          g^{c_1}(x + r_3\hat{\mu}_3 )  \ ,
\nonumber
\end{eqnarray}
where we apply the Lattice QCD prescription of linking the fields with links, to maintain the gauge 
invariance of our operator.  We also assume the sum over repeated indices.
The non-relativistic potential requires the computation of the Green functions 
present in eq. (\ref{Green function}).  Assuming that the Gluons are static,
and that moreover any permutation of gluons is left for the future application of 
the present static potential in constituent gluon models, the contraction of the 
gluon field operators provides adjoint temporal links, giving rise to the gluon operator,
\begin{eqnarray}
W_{3g}^A&= & f_{abc} \ f_{a'b'c'}
\label{glue888}
\\
&\Big[ &
 \widetilde U_{\mu_1} (x) \cdots  \widetilde U_{\mu_1} (x + (r_1-1)\hat{\mu}_1)
 \nonumber \\
&&
  \widetilde  U_4 (0,x + r_1 \hat{\mu}_1) \cdots 
   \widetilde U_4 (t-1,x+r_1\hat{\mu}_1) 
\nonumber \\
&&
    \widetilde U^\dagger_{\mu_1} (t,x + (r_1-1) \hat{\mu}_1) \cdots 
                  \widetilde U^\dagger_{\mu_1} (t,x  )\Big]^{a a'}
                 ~ \times
\nonumber \\
&\Big[ &
 \widetilde U_{\mu_2} (x) \cdots  \widetilde U_{\mu_2} (x + (r_2-1)\hat{\mu}_2)
 \nonumber \\
&&
  \widetilde  U_4 (0,x + r_2 \hat{\mu}_2) \cdots 
   \widetilde U_4 (t-1,x+r_2\hat{\mu}_2) 
\nonumber \\
&&
    \widetilde U^\dagger_{\mu_2} (t,x + (r_2-1) \hat{\mu}_2) \cdots 
                  \widetilde U^\dagger_{\mu_2} (t,x  )\Big]^{b b'}
                 ~ \times
\nonumber \\
&\Big[ &
 \widetilde U_{\mu_3} (x) \cdots  \widetilde U_{\mu_3} (x + (r_3-1)\hat{\mu}_3)
 \nonumber \\
&&
  \widetilde  U_4 (0,x + r_3 \hat{\mu}_3) \cdots 
   \widetilde U_4 (t-1,x+r_3\hat{\mu}_3) 
\nonumber \\
&&
    \widetilde U^\dagger_{\mu_3} (t,x + (r_3-1) \hat{\mu}_3) \cdots 
                  \widetilde U^\dagger_{\mu_3} (t,x  )\Big]^{c c'}
\ .
\nonumber 
\end{eqnarray}

\begin{figure}[t!]
\begin{picture}(350,220)(0,0)
\put(-30,100){\includegraphics[width=0.6\textwidth]{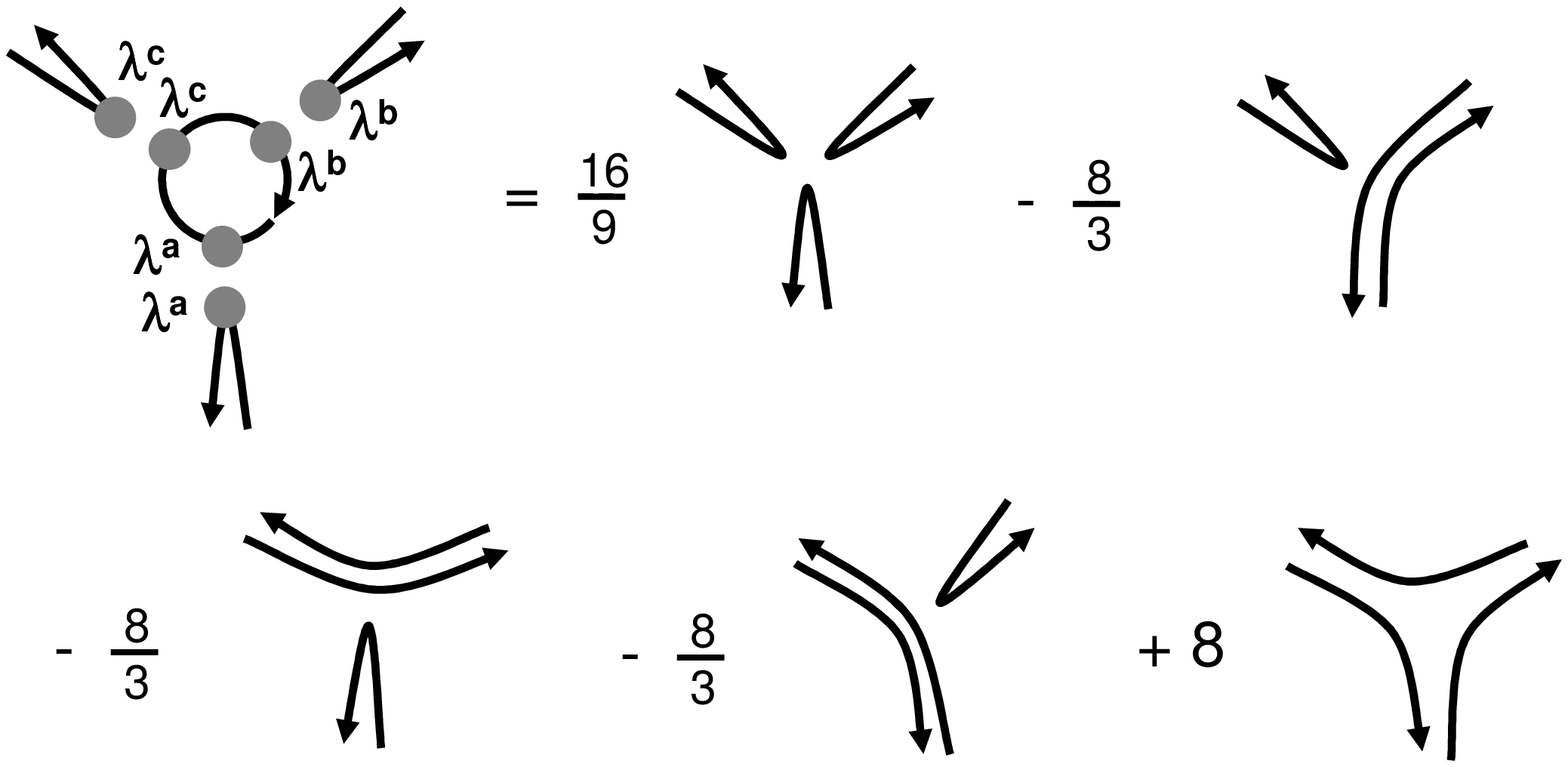}}
\put(-30,-25){\includegraphics[width=0.6\textwidth]{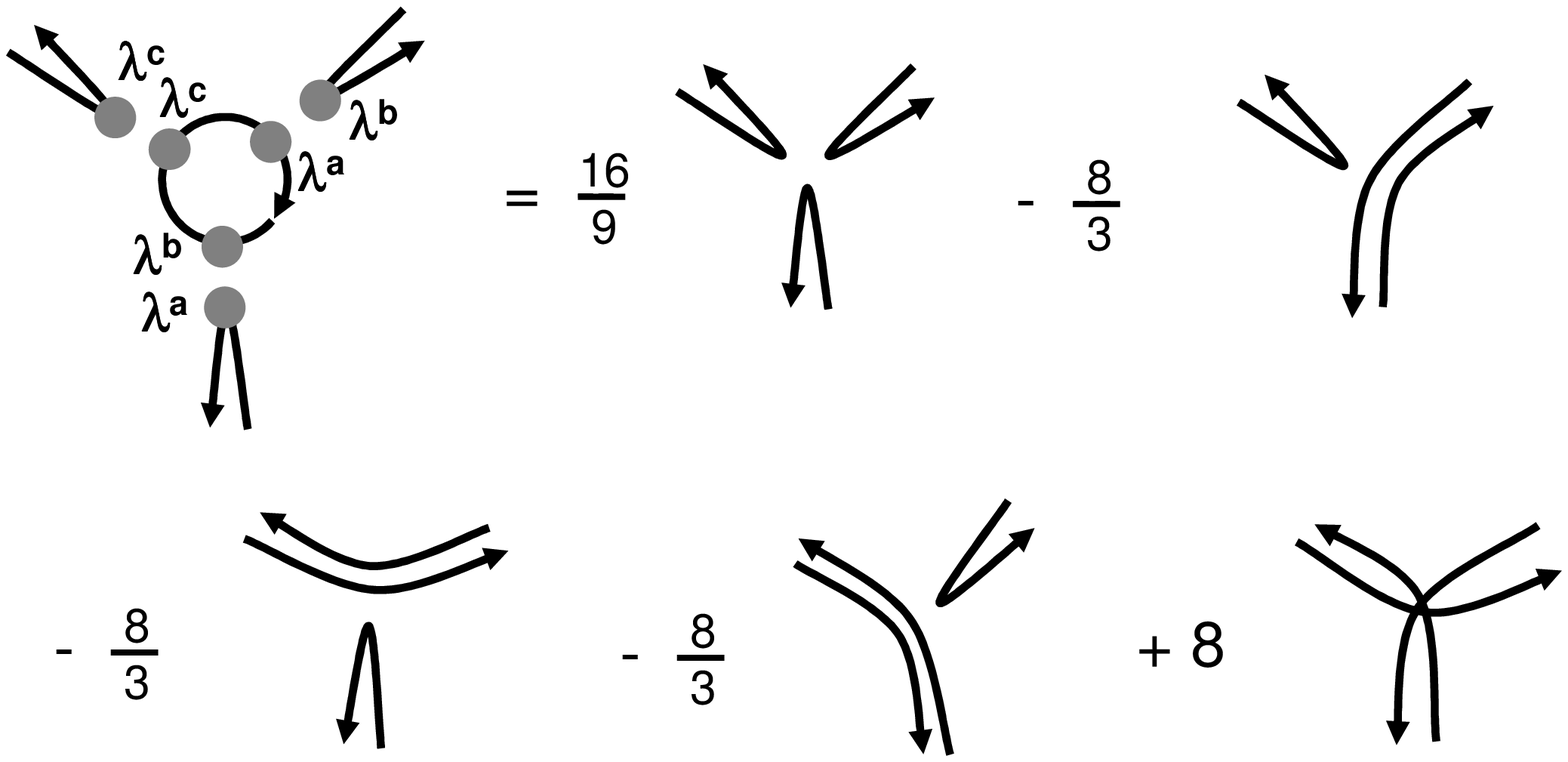}}
\end{picture}
\caption{Contractions of the three pairs of Gell-Mann matrices Graphical resulting from one of the
three-gluon wavefunctions. This shows that the three-gluon Wilson loops, present in
eqs. \eqref{asymwilson} and \eqref{symwilson}, are gauge invariant, because
they can be written as connected paths of Lattice QCD links $U$. }
\label{3contractions}
\end{figure}

\begin{figure}[t!]
\begin{picture}(350,180)(0,0)
\put(-30,-20){\includegraphics[width=0.6\textwidth]{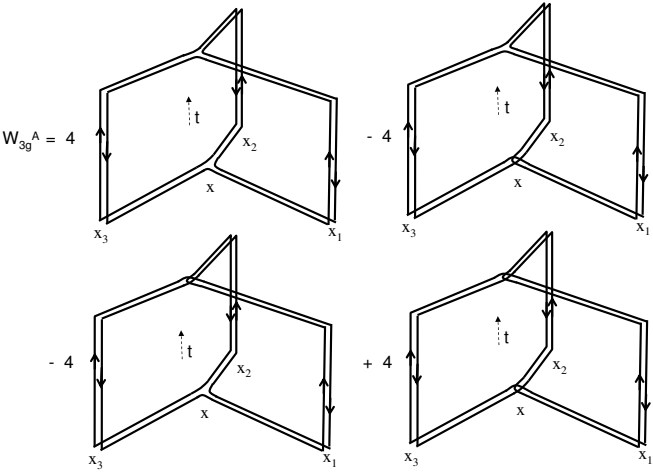}}
\end{picture}
\caption{The antisymmetric three-gluon Wilson loop $W_{3g}^A$ expressed with 
paths of quark-like fundamental $U$ links. }
\label{WilsonLoop3G_A}
\end{figure}

We now translate the adjoint links into quark links. This is convenient, both to explicitly
show that our Wilson loop is $SU(3)$ gauge invariant, and to arrive at a more convenient
expression for our computer simulations. 
So let us consider the product of two adjoint links, and apply
again the Fierz relation to, say,
\begin{eqnarray}
\sum_b	\tilde{U_1}^{ab}{ \tilde{U_2}}^{bc}
&=&
 \sum_b	
 \frac{1}{4} \mbox{Tr}[ {U_1}^{\dagger}\lambda^a U_1 \lambda^b  ]   \mbox{Tr}[ \lambda^b U_2 \lambda^c {U_2}^{\dagger} ]   
  \nonumber \\ &=&
 2 \frac{1}{4}  \mbox{Tr}[ {U_1}^{\dagger}\lambda^a U_1  U_2 \lambda^c {U_2}^{\dagger} ]   
   \nonumber \\  && \hspace{1cm}
- {2 \over 3}\frac{1}{4}  \mbox{Tr}[ {U_1}^{\dagger}\lambda^a U_1   ]  \mbox{Tr}[  U_2 \lambda^c {U_2}^{\dagger} ]   
 \nonumber \\ &=&
 \sum_b	\widetilde{U_1 U_2}^{ac}\ .
\end{eqnarray}
Thus the product of two adjoint links is the adjoint of the product of two links. Iterating this result to the product of 
an arbitrary number of links, we get that all three paths present in eq. (\ref{glue888}) verify,
\begin{eqnarray}
\Big[  \widetilde U_{\mu_1} (0, x) \cdots    \widetilde U^\dagger_{\mu_1} (t,x  )\Big]^{a a'}
&=&
{1 \over 2} \mbox{Tr} \Big\{ \lambda^a \, 
U_{\mu_1} (x) \cdots    U^\dagger_{\mu_1} (t,x  )
\nonumber \\ 
&& \  \  \  \lambda^{ a'}U_{\mu_1} (t,x) \cdots    U^\dagger_{\mu_1} (0,x  )
\Big\}
\nonumber \\
&=&
{1 \over 2} \mbox{Tr} \Big\{ \lambda^a \, 
X \lambda^{ a'}X^\dagger  \Big\}
\nonumber \\
&=& \widetilde X \ ,
\end{eqnarray}
where $X$ is the quark path utilized in the Wilson loop for static baryon potentials,
corresponding  to the gluon path $\widetilde X$ 
In particular,  the Wilson loop in eq. (\ref{glue888})
can be decomposed in quark paths $X$, $Y$ and $Z$, as in Fig.  \ref{Wilson loop diagram}, 
\begin{eqnarray}
W_{3g}^A&= & f_{abc} \ f_{a'b'c'}
\mbox{Tr} \Big\{  \lambda^a  X  \lambda^{a '}X^\dagger \Big\}
\nonumber \\
&&
\mbox{Tr} \Big\{  \lambda^b  Y  \lambda^{b '}Y^\dagger \Big\}
\mbox{Tr} \Big\{  \lambda^a  Z  \lambda^{c '}Z^\dagger \Big\} \ ,
\nonumber \\
W_{3g}^S&= & d_{abc} \ d_{a'b'c'}
\mbox{Tr} \Big\{  \lambda^a  X  \lambda^{a '}X^\dagger \Big\}
\label{asymwilson}
\\
&&
\mbox{Tr} \Big\{  \lambda^b  Y  \lambda^{b '}Y^\dagger \Big\}
\mbox{Tr} \Big\{  \lambda^a  Z  \lambda^{c '}Z^\dagger \Big\} \ ,
\label{symwilson} 
\end{eqnarray}
extending the three-quark Wilson loop of eq. (\ref{3qlooppaths}),
replacing the quark fundamental $SU(3)$ path $X$, 
by the gluon  adjoint $SU(3)$ path $\widetilde X$. 
We also removed the overall $1/8$ factors since the potentials are independent of the norm of the Wilson loops.

\begin{figure}[t!]
\begin{picture}(350,350)(0,0)
\put(-30,150){\includegraphics[width=0.6\textwidth]{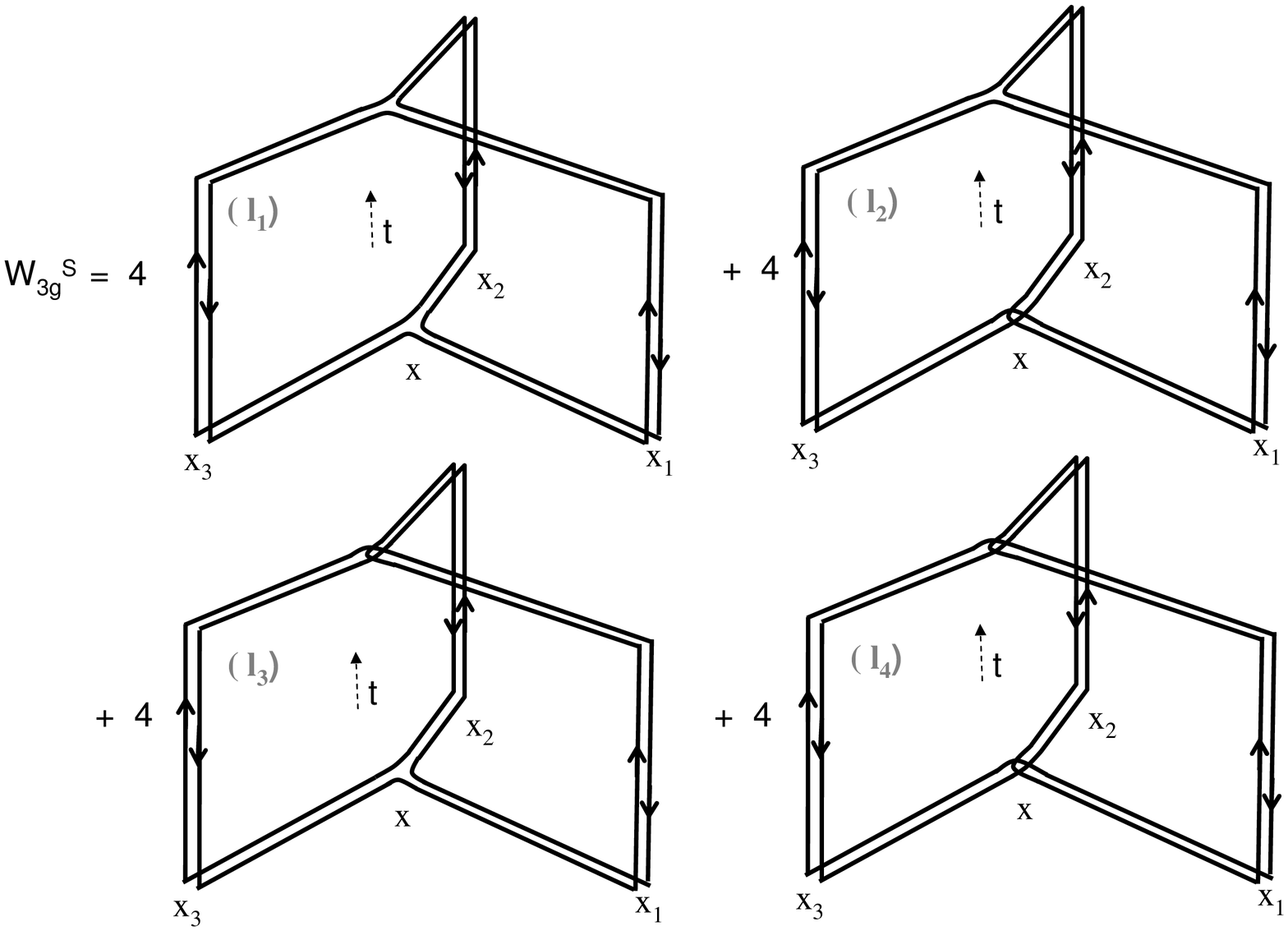}}
\put(-30,-20){\includegraphics[width=0.6\textwidth]{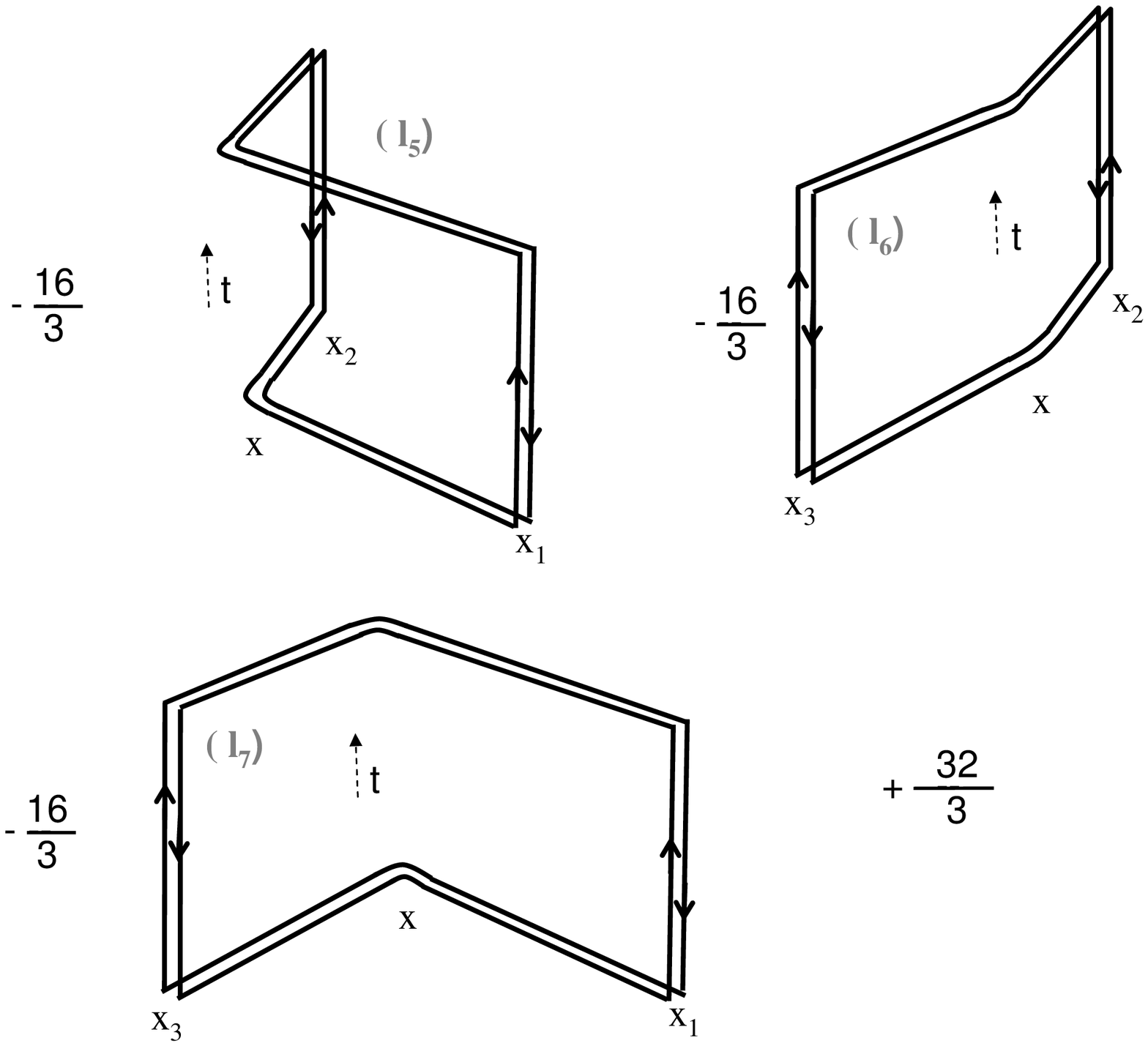}}
\end{picture}
\caption{The symmetric three-gluon Wilson loop $W_{3g}^S$ expressed with 
paths of quark-like fundamental $U$ links. Each individual loop is labeled by a $l_i$. }
\label{WilsonLoop3G_S}
\end{figure}

We now proceed to completely translate the results of eqs. \eqref{asymwilson} 
and \eqref{symwilson} into fundamental quark paths.  We express the 
eqs. \eqref{asymwilson} and \eqref{symwilson} in terms of 
correlations of the quark paths $X, \ Y, \ Z$  only. Noticing,
\begin{eqnarray}
f_{abc} &=& { 1 \over 4 \, i} \mbox{Tr} \left\{ \right(\lambda^a \lambda^b  - \lambda^b \lambda^a \left)\lambda^c  \right\} \   ,
\nonumber \\
d_{abc} &=& {1 \over 4} \mbox{Tr} \left\{\right(\lambda^a \lambda^b  + \lambda^b \lambda^a \left)\lambda^c  \right\} \   ,
\label{fdOfTrace}
\end{eqnarray}
we replace in eqs. \eqref{asymwilson} and \eqref{symwilson}the structure functions
 $f_{abc}$  and  $d_{abc}$ by traces of Gell-Mann matrices. Then we repeatedly apply the Fierz relation 
(\ref{Fierz relation}),  illustrated in Fig. \ref{FierzContraction} .

Subtracting and summing the results of the two different contractions of Fig.  \ref{3contractions}, 
we get the contribution of the respective symmetric and antisymmetric wavefunctions to the
three-gluon Wilson loops.  
\begin{eqnarray}
\label{f_cont1}
	&& f_{abc} Tr[ \lambda^a A ] Tr[ \lambda^b B ] Tr[ \lambda^c C ] = \\ \nonumber
	&& \hspace{.5cm} = \frac{i}{2} \lambda^a_{ij} \lambda^b_{kl} 
	\left( \lambda^b \lambda^a - \lambda^a \lambda^b \right)_{mn} A_{ji} B_{lk} C_{nm} \\ \nonumber
	&& \hspace{.5cm} = \frac{i}{2} \left( \lambda^a_{ij} \lambda^b_{kl} \lambda^b_{mp} \lambda^a_{pn} -
	\lambda^a_{ij} \lambda^b_{kl} \lambda^a_{mp} \lambda^b_{pn} \right) A_{ji} B_{lk} C_{nm}
\\ \nonumber	
  && \hspace{.5cm} = 2 i \left( Tr[ C B A ] - Tr[ A B C ] \right)\ ,
\end{eqnarray}
where we assumed a sum over repeated indices.
Following a similiar procedure we also get,
\begin{eqnarray}
	\label{f_cont2}
&&	f_{abc} Tr[ \lambda^a A \lambda^b B \lambda^c C ] = 
\\ \nonumber
&& \hspace{.5cm} = 2 i Tr[ A ] Tr[ B ] Tr[ C ] - 2 i Tr[ C B A ]
\end{eqnarray}
\begin{figure}[t!]
\begin{picture}(350,120)(0,0)
\put(-35,-20){\includegraphics[width=0.75\textwidth]{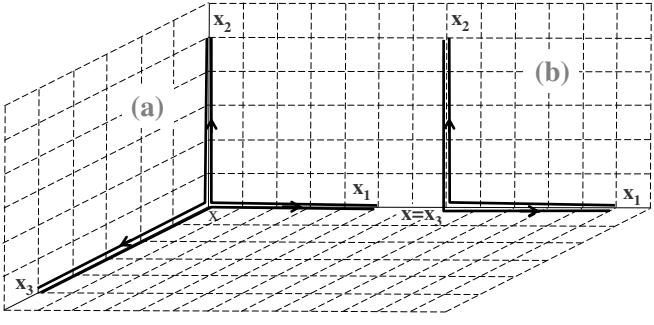}}
\end{picture}
\caption{Spatial paths, (a) for an equilateral triangle using the vertex of a cube, (b) 
for an isosceles rect triangle , using the vertex of a square. }
\label{spacepaths}
\end{figure}
Using the results \eqref{f_cont1} and \eqref{f_cont2} we finally arrive at the expression for the Wilson loop for the
antisymmetric colour arrangement
\begin{eqnarray}
\label{symmetricoperator}
	W_{3g}^{A} &=& + 4 Tr[ X Y^\dagger ] Tr[ Y Z^\dagger ] Tr[ Z X^\dagger ] 
	\\ \nonumber &&
		+ 4 Tr[ X^\dagger Y ] Tr[ Y^\dagger Z ] Tr[ Z^\dagger X ]
	\\ \nonumber &&
		- 4 Tr[ X Z^\dagger Y X^\dagger Z Y^\dagger ] - 4 Tr[ X Y^\dagger Z X^\dagger Y Z^\dagger ] \ ,
\end{eqnarray}
depicted in Fig. \ref{WilsonLoop3G_A}.
Using the same techniques for the operator for the symmetric colour wavefunction,
\begin{widetext}
\begin{eqnarray}
	&& d_{abc} Tr[ \lambda^a A ] Tr[ \lambda^b B ] Tr[ \lambda^c C ] = 
\nonumber
\\ \nonumber
	&& \hspace{.5cm} = 2 Tr[ A B C ] + 2 Tr[ C B A ] - \frac{4}{3} Tr[ A ] Tr[ B C ] - \frac{4}{3} Tr[ B ] Tr[ C A ]
	- \frac{4}{3} Tr[ C ] Tr[ A B ] + \frac{8}{9} Tr[ A ] Tr[ B ] Tr[ C ]
\ ,
\\ \nonumber
	&& d_{abc} Tr[ \lambda^a A \lambda^b B ] Tr[ \lambda^c C ] = 
\\ \nonumber
	&& \hspace{.5cm} = 2 Tr[ A C ] Tr[ B ] + 2 Tr[ B C ] Tr[ A ] + \frac{8}{9} Tr[ A B ] Tr[ C ]
	- \frac{4}{3} Tr[ A B C ] - \frac{4}{3} Tr[ C B A ] - \frac{4}{3} Tr[A] Tr[B] Tr[C] 
\ ,
\\  
	&& d_{abc} Tr[ \lambda^a A \lambda^b B \lambda^c C ] = 
\\  \nonumber
	&& \hspace{.5cm} = 2 Tr[ C B A ] + \frac{8}{9} Tr[ A B C ]
	- \frac{4}{3} Tr[ A ] Tr[ B C ] - \frac{4}{3} Tr[ B ] Tr[ C A ] - \frac{4}{3} Tr[ C ] Tr[ A B ]
	+ 2 Tr[A] Tr[B] Tr[C] 
\ , 
\end{eqnarray}
and finaly we get ,
\begin{eqnarray}
	W_{3g}^S &=& 4 Tr[ X Y^\dagger Z X^\dagger Y Z^\dagger ] + 4 Tr[ X^\dagger Z Y^\dagger X Z^\dagger Y ]
	- \frac{16}{3} Tr[ X Y^\dagger ] Tr[ X^\dagger Y ] - \frac{16}{3} Tr[ Y Z^\dagger ] Tr[ Y^\dagger Z ]
\nonumber \\ &&
	- \frac{16}{3} Tr[ Z X^\dagger ] Tr[ Z^\dagger X ]
	+ 4 Tr[ X^\dagger Y ] Tr[ Y^\dagger Z ] Tr[ Z^\dagger X ] + 4 Tr[ Y^\dagger X ] Tr[ Z^\dagger Y ] Tr[ X^\dagger Z ]
	+ \frac{32}{3} \ .
\label{antisymmetricoperator}
\end{eqnarray}
\end{widetext}
The results in terms of quark-like Wilson loops, composed of fundamental links only,
are illustrated in Figs.  \ref{WilsonLoop3G_A} and  \ref{WilsonLoop3G_S}.

\begin{figure}[t!]
\begin{picture}(350,180)(0,0)
\put(0,0){\includegraphics[width=0.45\textwidth]{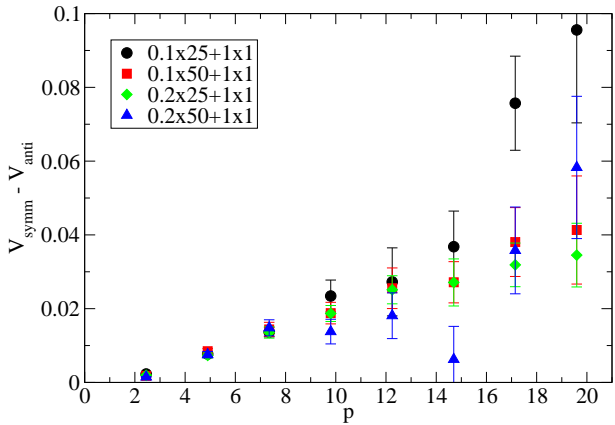}}
\end{picture}
\caption{We show the difference $V^S-V^A$  of the three-gluon potentials of the two operators $W_{3g}^S$ and $W_{3g}^A$ 
as a function of the perimeter $p$ of the respective triangle. 
For the spatial geometry of the loops, we utilize the equilateral  triangle of Fig. \ref{spacepaths}$(a)$.  
The different points plotted correspond to different smearings 
in space and to the same smearing of $1  \times 1$ in time. }
\label{differencepots}
\end{figure}

\section{Analytical discussion }

The class of Wilson loops $W_{3g}^A$ and $W_{3g}^S$ formally derived in Section II
still contain degrees of freedom, that we may use to increase the signal to noise ratio. In particular,
the paths linking the fixed positions $\mathbf x_1, \ \mathbf x_2$ and $\mathbf x_3$ of the three gluons,  
remain to be determined. 

Notice that smearing is a standard technique to increase the signal to noise ratio of the Wilson loop.  
The smearing 
\cite{sm1,sm2,sm3,sm4,sm5,sm6}
of the spatial links is a technique consisting in repeatedly mixing a link to neighbour 
staple-like paths. The resulting mixing is unitarized back to a $SU(3)$ matrix. 
The smearing is expected to maximize the signal (of the groundstate) to noise 
ratio when the smearing is comparable to the actual width of the QCD confining flux tube. Thus we
should try different smearings to arrive at the best signal of the groundstate, provided by the
exponential decay in eq. (\ref{Green function}). 

Moreover, the Wilson loops $W_{3g}^A$ and $W_{3g}^S$ defined in Section II  depend
on the position of the point $\mathbf x$, initially defined in eq. (\ref{mesonmesonmeson}).
Notice however that the actual static potential should not depend on this $\mathbf x$ point.
Possibly, as long as we keep fixed the points $\mathbf x_1, \ \mathbf x_2$ and $\mathbf x_3$,
the spatial paths connecting these points could also be arbitrarily changed,  even if they don't 
meet in a common point $\mathbf x$, however this remains  to be verified.  Importantly, we expect that 
the spatial paths closer to the actual position of the strings confining the three gluons will maximize the 
signal to noise ratio. 

In Fig.  \ref{spacepaths} we show two possible different spatial paths linking
the points $\mathbf x_1, \ \mathbf x_2$ and $\mathbf x_3$. In this paper, for simplicity, we
use only paths parallel to the lattice grid. In Fig. \ref{spacepaths}  $(a)$ we
place the three gluons at the vertices of an equilateral triangle, constructed with the egdes of a cube.
Placing the vertex of the cube at, say $(0,0,0)$, three points forming the triangle are,
$(r,0,0)$, $(0,r,0)$ and $(0,0,r)$. In Fig \ref{spacepaths}$(a)$, the $\mathbf x$ point  is
located at the simplest possible position for a numerical simulation, at the vertex $(0,0,0)$ of the cube. 
In Fig \ref{spacepaths}$(b)$, the paths are quite simple, we
place the three gluons at the vertices of an isosceles rect triangle, and the $\mathbf x$ point coincides
with the $\mathbf x_3$, thus the spatial geometry is planar. The paths in 
Fig. \ref{spacepaths}$(a)$  and $(b)$ are neither placed at the starfish-like string position, nor at the
position of the triangle-like string position. More sophisticated choices of paths might lead to better
signal to noise ratios, but the paths in Fig.  \ref{spacepaths} are the simplest for a first simulation.

\begin{figure}[t!]
\begin{picture}(350,180)(0,0)
\put(0,0){\includegraphics[width=0.45\textwidth]{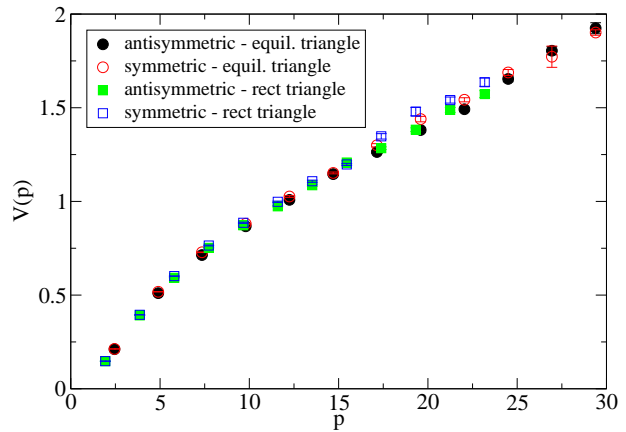}}
\end{picture}
\caption{We show the three-gluon potentials for the two operators( $W_{3g}^A$ and $W_{3g}^S$ )
as a function of the perimeter $p$ of the respective triangle.
We utilize both the equilateral  triangle and the isosceles rect triangle spatial paths
of Fig. \ref{spacepaths}$(a)$  and $(b)$. The results are extracted from 141 
$SU(3)$  Lattice QCD configurations $24^3 \times 48$, with the smearing 
of $ 50 \times  0.2$  in space and of $1  \times 1$ in time. }
\label{firstpots}
\end{figure}
 
On the other hand we may explore analytical similarities or differences
between the Wilson loops $W_{3g}^A$ and $W_{3g}^S$.
The Casimir scaling, dominating the pertubative QCD,  and, at least,
the short distance  potentials, can be algebraicly computed, 
\begin{equation}
\lambda_1 \cdot \lambda_2 = 
{
 \left( \lambda_1 + \lambda_2 + \lambda_3 \right)^2 
- \left( {\lambda_1}^2 + { \lambda_2 }^2  + { \lambda_3 }^2 \right)
\over 6 }= - 6 \ .
\end{equation}
and the result is the same both for the symmetric and the antisymmetric potentials. 
Thus the short range part of the interactions should be identical.

Now we also check that in the limit where two gluons are superposed, we recover
the normal two-gluon operator, where the result is
proportional to (the proportionality factor is irrelevant here),
\begin{equation}
W_{gg}=W_{q \bar q}\  {  W_{q \bar q}} ^* - 1
\end{equation} 
where, say,  $W_{q \bar q}= $Tr$ \left\{X \, Y^\dagger \right\}$ is a complete one-quark Wilson Loop. 
to decay exponentially with large times. Thus when $x_3=x_2$ or equivalently when
$Z=Y$, we get, for the antisymmetric loop  $W_{3g}^A$,
\begin{eqnarray}
	W_{3g}^{A} &\rightarrow &  24 \left( W \, W^* -1 \right)
\end{eqnarray}
and are also identical in the symmetric loop  $W_{3g}^S$,
\begin{eqnarray}
	W_{3g}^S &\rightarrow &   {40 \over 3} \left( W \, W^* -1 \right) \ .
\end{eqnarray}
Importantly, since the result only differs in a physically irrelevant 
constant factor, this shows that whenever two of the arms of the starfish are superposed, 
the two potentials, for the symmetric and for the antisymmetric cases are identical.
Then, if any difference occurs, it only occurs when the arms are separated. Thus we should
position the gluons at the vertices of an open triangle, say an equilateral triangle, or an isosceles 
rect triangle, to study this possible difference.

\section{Numerical results }

Since this is mainly an analytical paper, in Section
IV we only numerically simulate the simplest paths to compute,
with the spatial sub-paths depicted in Fig. \ref{spacepaths} . 
We perform our simulations with 141 configurations generated by the
Monte Carlo method in a $24^3 \times 48$ periodic Lattice, with $\beta=6.2$ 
and $a \sim 0.072$ fm.

\begin{figure}*[t!]
\includegraphics[width=0.5\textwidth]{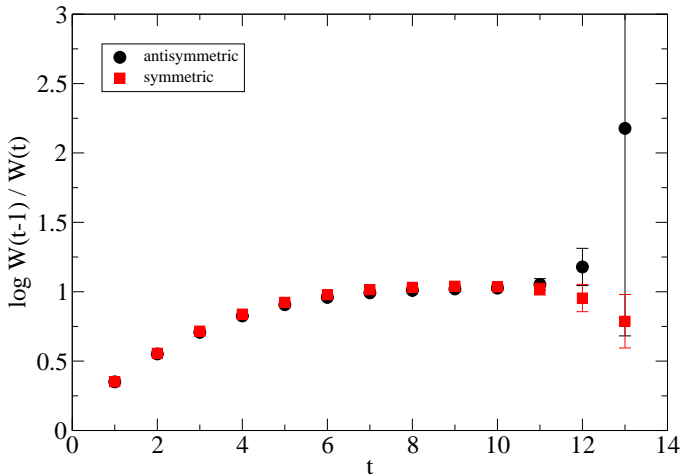}
\caption{Effective mass plots, 
for the two operators( $W_{3g}^A$ and $W_{3g}^S$ ),
for the equilateral triangle geometry with a perimeter $p=15 \sqrt{2}$ and for 141
Lattice QCD configurations $24^3 \times 48$.  In this effective mass plot we use 
$ 50 \times  0.2$  smearing steps in space and $1  \times 1$ smearing step in time.
 }
\label{massplots}
\end{figure}

First we check that the sum of all the different quark-like Wilson loops vanish
in the limit of large euclidian time $t$. This actually happens, and we also 
numerically check that the Wilson loops , as described in Figs.  \ref{WilsonLoop3G_A}
and \ref{WilsonLoop3G_S}, in the limit of large $t$ tend to,
\begin{eqnarray}
l_1 =l_2=l_3=l_4& \rightarrow { 1 \over 3}
\nonumber \\
\nonumber \\
l_5=l_6=l_7 & \rightarrow 1 \ ,
\end{eqnarray}

Then, we study the possible difference between the antisymmetric and
symmetric static potentials, defined in Figs.  \ref{WilsonLoop3G_A}
and \ref{WilsonLoop3G_S} and in eqs. (\ref{symmetricoperator}) 
and (\ref{antisymmetricoperator}). Notice that we explore different
smearings of the spatial paths, because our spatial paths neither coincide
with the vortex positions of the starfish-like model, nor with the vortex positions 
of the starfish-like model. Nevertheless the resulting difference, although
small, shows little dependence on the smearing. The results of our simulations
for the  difference between the antisymmetric and symmetric static potentials 
is show in Fig. \ref{differencepots},  suggesting a difference
\begin{equation}
V_{sym} -V_{asym}\simeq   0.04 \sigma \, p \ ,
\end{equation}
where $p= r_{12}+r_{23}+r_{31}$ is the perimeter of our equilateral triangle 
and the sum of the three inter-gluon distances, and $\sigma$ is the string tension of the 
fundamental quark-antiquark potential.
Both the small difference and its linear behaviour confirm our analytical study of Section III,
where we showed that for short distances the difference $V_{sym} -V_{asym}$ is 
vanishing.

To verify that the static potentials do not depend on the arbitrary meeting point
$\mathbf x$ of the spatial paths, we compute the static potentials for two different
geometries, depicted in Fig. \ref{spacepaths}. In case $(a)$ ,
the  point $\mathbf x$  is placed relatively far from the position of any of the three gluons.
In case $(b)$ the  point $\mathbf x$  coincides with the position of one of the gluons.
As anticipated in Section III, the potentials show little dependence on the point $\mathbf x$.
This is illustrated in Fig. \ref{firstpots}, where both geometries produce similar results.

We also study the absolute size of the antisymmetric and
symmetric static potentials, defined in Figs.  \ref{WilsonLoop3G_A}
and \ref{WilsonLoop3G_S} and in eqs. (\ref{symmetricoperator}) 
and (\ref{antisymmetricoperator}).  Again, we explore different
smearings of the spatial paths.  It occurs that the absolute value
of the potentials are more smearing-dependent that the nearly
smearing-independent difference of the potentials. In particular
we cannot yet establish precisely the strength of the coulomb potential.
Nevertheless the results of our simulations show in Fig. \ref{firstpots},  
suggest that both the antisymmetric and the symmetric static potentials are close 
to the triangle-like model,
\begin{equation}
V_{\mbox{triangle}} \simeq   \sum_{i < j} - {\alpha \over  r_{ij}}+  \sigma  \, p,
\end{equation}
where $p$ is the perimeter of the triangle.
Our results are clearly not compatible with the starfish-like model. In the equilateral triangle 
spatial geometry, the starfish-like model corresponds to a  linear component of the potential 
component 30 \% larger than the one of the triangle-like model. The starfish-like a linear term
is ${9 \over 4}\sigma  \,  l_{min}$, where $l_{min}$ is the sum of the distances of the 
gluons to the Fermat -Torricelli point, minimizing the total distance of the adjoint strings in the 
starfish-like model.

Finally, to check that the $24^3 \times 48$ Lattice configurations
are producing good results, we show the effective mass plot of
a static potential in Fig.  \ref{massplots}.

To conclude, we show that there are two, and only two, symmetric and antisymmetric, three-gluon 
static potential. We derive the two respective Wilson loops and study them analytically. We perform 
numerical tests, verifying that our Wilson loop is correct. Notice that the three-gluon 
Wilson loops include products of up to three fundamental Wilson loops, technically difficult to compute.
We thus leave the systematic numerical exploration of the three-gluon Wilson loops for future works.
Nevertheless, our numerical simulations already indicate that the symmetric potential is slightly larger 
than the antisymmetric one, and that both are compatible with the triangle-like model for the three-gluon 
static potential.

\acknowledgments

Part of the present work was funded by the FCT grants 
PDCT/FP/63923/2005 and POCI/FP/81933/2007.
We thank Orlando Oliveira for sharing with us his set of 141
 configurations in a $24^3 \times 48$ SU(3)QCD Lattice
in part based on the MILC collaboration's public Lattice gauge
theory code
\cite{Gauge,Blum},
 and for scientific discussions.


%

\begin{thebibliography}{99}
%

\bibitem{LlanesEstrada:2000jw}
  F.~J.~Llanes-Estrada, S.~R.~Cotanch, P.~J.~de A. Bicudo, J.~E.~F.~Ribeiro and A.~P.~Szczepaniak,
  Nucl.\ Phys.\  A {\bf 710}, 45 (2002)
  [arXiv:hep-ph/0008212].
%
\bibitem{Meyer:2004jc}
  H.~B.~Meyer and M.~J.~Teper,
  Phys.\ Lett.\  B {\bf 605}, 344 (2005)
  [arXiv:hep-ph/0409183].
%
\bibitem{LlanesEstrada:2005jf}
  F.~J.~Llanes-Estrada, P.~Bicudo and S.~R.~Cotanch,
  Phys.\ Rev.\ Lett.\  {\bf 96}, 081601 (2006)
  [arXiv:hep-ph/0507205].
  
  
  \bibitem{Hou:1982dy}
  W.~S.~Hou and A.~Soni,
  Phys.\ Rev.\  D {\bf 29}, 101 (1984).


\bibitem{Mathieu:2006bp}
  V.~Mathieu, C.~Semay and B.~Silvestre-Brac,
  Phys.\ Rev.\  D {\bf 74}, 054002 (2006)
  [arXiv:hep-ph/0605205].

\bibitem{Buisseret:2007hf}
  F.~Buisseret and C.~Semay,
  Phys.\ Rev.\  D {\bf 76}, 017501 (2007)
  [arXiv:0704.1753 [hep-ph]].

\bibitem{Mathieu:2008pb}
  V.~Mathieu, C.~Semay and B.~Silvestre-Brac,
  Phys.\ Rev.\  D {\bf 77}, 094009 (2008)
  [arXiv:0803.0815 [hep-ph]].

\bibitem{Boulanger:2008aj}
  N.~Boulanger, F.~Buisseret, V.~Mathieu and C.~Semay,
  arXiv:0806.3174 [hep-ph].


%
\bibitem{Leinweber:1998uu}
  D.~B.~Leinweber, J.~I.~Skullerud, A.~G.~Williams and C.~Parrinello  [UKQCD
                  Collaboration],
  Phys.\ Rev.\  D {\bf 60}, 094507 (1999)
  [Erratum-ibid.\  D {\bf 61}, 079901 (2000)]
  [arXiv:hep-lat/9811027].


\bibitem{Oliveira_propagator}
  P.~J.~Silva and O.~Oliveira,
  Nucl.\ Phys.\ B {\bf 690}, 177 (2004)
  [arXiv:hep-lat/0403026].

\bibitem{Fischer:2002eq}
  C.~S.~Fischer, R.~Alkofer and H.~Reinhardt,
  Phys.\ Rev.\  D {\bf 65}, 094008 (2002)
  [arXiv:hep-ph/0202195].
  
\bibitem{Nielsen:1973cs}
  H.~B.~Nielsen and P.~Olesen,
  Nucl.\ Phys.\ B {\bf 61} (1973) 45.
  
  
\bibitem{Michael:1985ne}
  C.~Michael,
  Nucl.\ Phys.\  B {\bf 259}, 58 (1985).

  
\bibitem{Griffiths:1983ah}
  L.~A.~Griffiths, C.~Michael and P.~E.~L.~Rakow,
  Phys.\ Lett.\  B {\bf 129}, 351 (1983).


\bibitem{Szczepaniak:1995cw}
  A.~Szczepaniak, E.~S.~Swanson, C.~R.~Ji and S.~R.~Cotanch,
  Phys.\ Rev.\ Lett.\  {\bf 76}, 2011 (1996)
  [arXiv:hep-ph/9511422].
\bibitem{Abreu:2005uw}
  E.~Abreu and P.~Bicudo,
  J.\ Phys.\ G {\bf 34}, 195207 (2007)
  [arXiv:hep-ph/0508281].





\bibitem{glulat1} C. J. Morningstar and M. Peardon, Phys. Rev. D {\bf 60},
034509
(1999).

\bibitem{glulat2} UKQCD Collaboration, G. S. Bali {\it et al.},
Phys. Lett. {\bf B309}, 378 (1993).

\bibitem{glulat3} H. Chen, J. Sexton, A. Vaccarino, and D. Weingarten,
Nucl. Phys. {\bf B34}, 357 (1994).

\bibitem{glulat4} M. Teper, hep-th/9812187 (1998).



\bibitem{Campbell:1985kp}
  N.~A.~Campbell, I.~H.~Jorysz and C.~Michael,
  Phys.\ Lett.\  B {\bf 167}, 91 (1986).

\bibitem{Bali:2000un}
  G.~S.~Bali,
  Phys.\ Rev.\  D {\bf 62}, 114503 (2000)
  [arXiv:hep-lat/0006022].

\bibitem{Okiharu:2004ve}
  F.~Okiharu, H.~Suganuma and T.~T.~Takahashi,
  Phys.\ Rev.\  D {\bf 72}, 014505 (2005)
  [arXiv:hep-lat/0412012].
\bibitem{Okiharu:2004wy}
  F.~Okiharu, H.~Suganuma and T.~T.~Takahashi,
  Phys.\ Rev.\ Lett.\  {\bf 94}, 192001 (2005)
  [arXiv:hep-lat/0407001].
  
  
\bibitem{Bicudo:2007xp}
  P.~Bicudo, M.~Cardoso and O.~Oliveira,
  Phys.\ Rev.\  D {\bf 77}, 091504 (2008)
  [arXiv:0704.2156 [hep-lat]].
\bibitem{Cardoso:2007dc}
  M.~Cardoso, P.~Bicudo and O.~Oliveira,
  PoS {\bf Lattice2007}, 293 (2007)
  [arXiv:0710.1762 [hep-lat]].
  
  

\bibitem{Jahn:2003uz}
  O.~Jahn and P.~de Forcrand,
  Nucl.\ Phys.\ Proc.\ Suppl.\  {\bf 129}, 700 (2004)
  [arXiv:hep-lat/0309115].

\bibitem{deForcrand:2005vv}
  Ph.~de Forcrand and O.~Jahn,
  Nucl.\ Phys.\  A {\bf 755}, 475 (2005)
  [arXiv:hep-ph/0502039].
  
  
\bibitem{Takahashi:2002it}
  T.~T.~Takahashi and H.~Suganuma,
  Phys.\ Rev.\ Lett.\  {\bf 90}, 182001 (2003)
  [arXiv:hep-lat/0210024].
  
\bibitem{Suganuma:2003cv}
  H.~Suganuma, T.~T.~Takahashi and H.~Ichie,
  arXiv:hep-lat/0312031.

\bibitem{Suganuma:2004hu}
  H.~Suganuma, H.~Ichie and T.~T.~Takahashi,
  arXiv:hep-lat/0407011.

\bibitem{Takahashi:2004rw}
  T.~T.~Takahashi and H.~Suganuma,
  Phys.\ Rev.\  D {\bf 70}, 074506 (2004)
  [arXiv:hep-lat/0409105].

\bibitem{Yamamoto:2007nn}
  A.~Yamamoto and H.~Suganuma,
  Phys.\ Rev.\  D {\bf 77}, 014036 (2008)
  [arXiv:0709.0171 [hep-ph]].

\bibitem{Suganuma:2008ej}
  H.~Suganuma, A.~Yamamoto, N.~Sakumichi, T.~T.~Takahashi, H.~Iida and F.~Okiharu,
  arXiv:0802.3500 [hep-ph].

\bibitem{Yamamoto:2008fm}
  A.~Yamamoto, H.~Suganuma and H.~Iida,
  arXiv:0805.4735 [hep-ph].

\bibitem{SilvestreBrac:2003sa}
  B.~Silvestre-Brac, C.~Semay, I.~M.~Narodetskii and A.~I.~Veselov,
  Eur.\ Phys.\ J.\  C {\bf 32}, 385 (2003)
  [arXiv:hep-ph/0309247].


  
  
\bibitem{bookGennes} 
P.G. de Gennes, {\it Superconductivity of Metals and Alloys}
(Addison-Wesley, Reading, MA, 1989).
    
\bibitem{sm1}
G. S. Bali, Schlichter and K. Schiling, Phys. Rev. D 51, 5165 ( 1995 )

\bibitem{sm2}
G. Parisi, R. Petronzio, and F. Rapuano, Phys. Lett.. B 128, 418 ( 1983 )

\bibitem{sm3}
APE Collaboration ( M. Albanese et al ), Phys. Lett. B 192, 163 ( 1987 )

\bibitem{sm4}
F. Okiharu and R. M. Woloshyn, Eur, Phys. J. C35, 537 (2004)

\bibitem{sm5}
T. T. Takahashi, H. Suganuma, Y. Nemoto and H. Matsufuru, Phys. Rev. D 65, 114509 (2002)

\bibitem{sm6}
G. S. Bali, H. Neff, T. D�ssel, T. Lippert and K. Schilling, Phys. Rev. D 71, 114513 (2005)



\bibitem{Gauge}
This work was in part based on the MILC collaboration's public Lattice gauge
theory code. See http://physics.indiana.edu/$\sim$sg/milc.html.
%
\bibitem{Blum}
T. Blum, C. DeTar, S. Gottlieb, K. Rummukainen,
Urs M. Heller, J. E. Hetrick, D. Toussaint,
R. L. Sugar, M. Wingate,
Phys. Rev. \textbf{D55}, R1133 (1997).

\end{thebibliography}
\end{document}